\documentclass[12pt]{article}
\RequirePackage[OT1]{fontenc}
\RequirePackage{amsthm,amsmath}
\RequirePackage[numbers]{natbib}
\usepackage{setspace}

\numberwithin{equation}{section}
\theoremstyle{plain}

\newcommand{\rr}{{ \Bbb R}}

\newcommand{\one}{{\hbox{1{\kern -0.35em}1}}}

\newcommand{ \beq}[1]{  \begin{eqnarray} \label{#1}}
\newcommand{\eeq}{\end{eqnarray}}
\newcommand{ \bed}{  \begin{displaymath}}
\newcommand{\eed}{\end{displaymath}}
\newcommand{ \bea}{\bed \begin{array}{rl}}
\newcommand{\eea}{\end{array}\eed}
\newcommand{\disp}{\displaystyle}
\newcommand{\al}{\alpha}\newcommand{\e}{\varepsilon}

\newtheorem{thm}{Theorem}[section]

\newtheorem{rem}[thm]{Remark}
\newtheorem{cor}[thm]{Corollary}
\newtheorem{defn}[thm]{Definition}

\newcommand{\thmref}[1]{Theorem~{\rm \ref{#1}}}

\def\openbox{$\sqcup\llap{$\sqcap$}$}
\def\endproof{\unskip \enskip
    \null \nobreak \hfill \openbox \par}

\newcommand{\ad}{&\!\!\!\disp}

\newcommand{ \barray}{ \b \begin{array}{ll}}
\newcommand{\earray}{\end{array}}

\usepackage{makeidx}  
\usepackage{amsfonts}
\makeindex            
\usepackage{times}  
\usepackage{graphicx} 

\date{}

\title{Optimal Extraction and Taxation of Strategic Natural Resources:\\
A Differential Game Approach }
\author{Moustapha Pemy\thanks{Department of Mathematics, Towson
University, Towson, MD 21252-0001, mpemy@towson.edu}}

\begin{document}

\maketitle

  \begin{abstract} 
This paper studies the optimal extraction and taxation of nonrenewable natural resources. It  is well known that  the market values of the main strategic resources such as oil, natural gas, uranium, copper,..., etc, fluctuate randomly following global and seasonal macroeconomic parameters, these values are modeled using Markov switching L\'evy processes. We formulate this problem as a differential game. The two players  of this differential game are the mining company whose aim is to maximize the revenues generated from its extracting activities and the government agency in charge of regulating and taxing natural resources. We prove the existence of a Nash equilibrium. The corresponding Hamilton Jacobi Isaacs equations are completely solved and the value functions as well as the optimal extraction and taxation rates are derived in closed-form. A Numerical example is presented to illustrate our findings.  
 
\end{abstract}

\vspace*{0.2in}


\section{Introduction}
Natural resources have always been the main sources of income for some  developing countries. The extraction of natural resources in developing countries is usually done by  multinational corporations. The revenues generated from the sales of these resources in world markets as well as the taxes  those countries levy on multinational  mining companies accounted for more than half of the budget of those  resource-rich developing countries. Thus, the production and regulation of strategic natural resources  have always been one of the prime topics of discussion in political and scientific circles.\\
 The earliest scientific contribution on the extraction of natural resources was obtained in the thirties by Hotelling \cite{Hotelling}, he derived  an optimal extraction policy under the assumption that the commodity price is constant. A wide range of  economists have extended the Hotelling model by taking into account the uncertainty when modeling commodity prices, for instance  Gibson and Schwartz \cite{GibSwc},  Schwartz \cite{Schw} and Cortazar  et al. \cite{CortSch} used stochastic mean reverting models for commodity prices.  One can cite the work of Pindyck \cite{Pindyck, Pindyck2},  Hanson \cite{Hanson2},  Lin and Wagner \cite{LinWagner},   Gaudet \cite{Gaudet},  Luo and Zhao \cite{DongkunZhao} for various extensions of the basic Hotelling model.
 Cherian  et al. \cite{Cherian} studied the optimal extraction of nonrenewable resources as a stochastic optimal control problem with two state variables, the commodity price and the size of the remaining reserve. They solved the control problem numerically by using Markov chain approximation methods.
 Recently Aleksandrov  et al. \cite{Alek} studied the optimal production of oil as an American-style real option and used Monte-Carlo methods to approximate the optimal production rate when the oil price follows a mean-reverting process.\\
The taxation of natural resources has also generated a great deal of interest in the academic literature. One can cite the work of Beals  et al. \cite{Beals} on tax and investment policies  for  hard minerals  and the contributions of Heaps and Helliwell \cite{Heaps}, and Bhattacharyya \cite{bba} on efficient tax policies for natural resources and energy,  Lin and Prince \cite{LinPrince} on optimal gas tax in California. \\
The extraction and taxation of natural resources are in fact two sides of the same coin when it comes to generating revenues for the public finances of resource-rich developing countries. However, throughout the scientific literature these issues have usually been treated separately. The main contribution of this work is that we treat these problems in their natural setting by highlighting the interplay between  extraction and taxation policies for strategic resources. 
 We use the framework of noncooperative differential games to tackle these issues. We formulate this problem as a differential game where the two players are the multinational mining company and the government.
 Obviously the multinational  company wants to maximize its share of profits from the sales of the extracted resource in commodity exchange markets  and the government also wants to maximize its share of profits from the sales of the extracted resource as well as the income tax it levies on the multinational company. Pemy \cite{pem4} studied the extraction and taxation of oil using mean-reverting regime switching L\'{e}vy processes to capture global disruptions in oil market and used the viscosity solution framework to estimate the optimal extraction and taxation rates.
  It is also self evident that commodity prices in exchange markets  fluctuate following various macro-economical and global  geopolitical forces. It is therefore important to take into account the random dynamic of the commodity value when solving the optimal extraction and  taxation problems.\\ 
In this paper, we use  regime switching L\'{e}vy processes to model commodity prices.  Regime switching models have been extensively used in the financial economics literature since their introduction by Hamilton \cite{Hamilton}.  Many 
authors have studied the control of systems that involve regime switching
 using a hidden Markov chain and/or jumps diffusions, one can cite  $\O$ksendal and Sulem \cite{oksendal}, Biswas et al. \cite{Biswas}, Davis et al. \cite{Davis}, Bayraktar et al. \cite{BayMen},   Pemy \cite{pem2, pem3, pem5} among others. \\ 
 In this of this work, we study the evolution of a production sharing agreement between a multinational mining company and the government of a resource-rich country using the framework of stochastic noncoorperative differential game. We prove the existence of a Nash equilibrium and derive the value functions as well as the optimal extraction and taxation policies in closed-form. 
The paper is organized as follows. In the next section, we
formulate the problem under consideration. In Section 3, we proof the existence of the Nash equilibrium. In section 4, we derive  the value functions and the optimal strategies.  A Finally, in section 5,  we present  a  numerical example that illustrates our results.

\section{Problem formulation}
Consider a company which has  a  Production Sharing Agreement with the government  of a country rich in natural resources. The agreement is for the extraction of a nonrenewable natural resource. Both parties will share the profits from the sales of the mineral on world markets following a simple rule where the company takes $100\theta$ percent and the government $100(1-\theta)$ percent of the profits, for some $\theta \in (0,1)$. We assume that the mining lease is a long term contract, thus we use the infinite time horizon framework to model this long term contract. Let $X_t$ be  the value of nonrenewable resource at time $t$. Given that commodity values are very sensitive to global macro-economical and geopolitical shocks, we model $X_t$ as a regime switching L\'{e}vy process with two states. Let $\al(t)\in{\cal M}=\{1,...,m\}$ be a finite state Markov chain that captures the state of the commodity marketplace.
Let  $(\eta_t)_t$ be a L\'{e}vy process and let $N$ be the Poisson random measure of $(\eta_t)_t$, $  N(t,U)=\sum_{0<s\leq t}{\bf 1}_U(\eta_s-\eta_{s^-})$  for any Borel set $U\subset \rr$. Moreover, let $\nu$ be the L\'{e}vy measure of $(\eta_t)_t$ we have $\nu(U)=E[N(1,U)]$ for any Borel set $U\subset \rr$. The differential form of $N$ is denoted by $N(dt,dz)$, we define the  differential $ \bar{N}(dt,dz)$ as follows
  \bea
\disp
  \bar{N}(dt,dz)=\left\{  \begin{array}{ll} N(dt,dz)-\nu(dz)dt \qquad &\hbox{if  } |z|<1\\
   N(dt,dz)&\hbox{if  } |z|\geq 1.   \end{array} \right.
   \eea
  We model the evolution  of the profit sharing agreement as a differential game where the two players are the mining company and the government. Each player acting as a controller will to maximize its own profit throughout the duration of the contract.  The mining company will try to maximize its share of profits from the sales of  mineral in world markets, while the government will also try to maximize both its share of the profits  from the sales of  mineral in world markets  and the income tax it levies on the mining company.   We denote the mining company as Player 1 and the government as Player 2.
 We  assume that  the process $ X(t)$ follows the dynamical system
  \beq{model}
 \left \{   \begin{array}{ll}
 &\disp dX(t)=[ X(t)\mu(\alpha(t)) -\rho u_1]\mathrm{d}t \disp +\sigma(\alpha(t))  X(t)dW(t)  \\
 & \hspace{0.5in}\disp +  X(t)\int_{\rr}\gamma(\alpha(t))z  \bar{N}(dt,dz) \bigg),\\
&\disp X(0)=x\geq 0,  \quad 0\leq t\leq \infty, \end{array} \right.
\eeq   
 where  $\rho\in[0,1)$ captures the relative impact of the extracting activities, $ u_1(t)\in U_1=[0,  \bar{u}_1]$ is the extraction rate chosen by the company and $ u_2(t) \in U_2= [\underbar{u}_2,  \bar{u}_2]$ is the tax rate chosen by the government.  Note that $  \bar{u}_2$ can be seen as the maximum tax rate that can be levied on any  company, this value is assumed to be known by both players, it comes naturally that $  \bar{u}_2<1$. The main idea we want to capture with the control $u_t(\cdot)$ is that the government may decide to give some tax breaks or exonerations to mining companies because of various macro-economical or political reasons in that case $u_2(t)$ represents the effective tax rate imposed to the mining company at time $t$.  It is also common that governments sometimes give tax subsidies to various industries for a wide range of political reasons, in order to capture that other possibility, the lowest possible tax rate that can be imposed to the mining company $\underbar{u}_2$ can take negative values, $-1<\underbar{u}_2\leq 0$.    We assume that for each state $i\in {\cal M}$, $\mu(i), \sigma(i)$ and $\gamma(i)$ are known constants, they respectively represent the rate of return, volatility and jump intensity.  The processes $ u_1(t)$ and $ u_2(t)$  are control variables, and  $ W(t)$ is the Wiener process  defined on a probability space  $ (\Omega,{\cal F },P)$. Moreover, we assume that $W(t)$, $\eta_t$ and $\al(t)$ are independent.
The generator of this Markov chain is $
Q=(q_{ij})_{ij}$, with $q_{ij}\geq 0, i\neq j$, $q_{ii}=-\sum_{j\neq i}q_{ij}$.
  \begin{rem}
Our commodity pricing model (\ref{model}) covers two main possibilities.
  \begin{enumerate}
\item If the country is a small or medium size producer, then the extraction rate will not affect the global price of the commodity, thus $\rho = 0$  and  the dynamic of $X_t$ becomes
  \beq{mpdel1}
dX(t)&= X(t) \bigg(\mu(\alpha(t))\mathrm{d}t \disp +\sigma(\alpha(t)) dW(t)  \nonumber\\
 & \hspace{0.5in}\disp + \int_{\rr}\gamma(\alpha(t))z  \bar{N}(dt,dz) \bigg),
\eeq  
\item If the country is a major producer that really influences the global price of the commodity such as Saudi Arabia is for oil, then the  parameters $\rho>0$. So the extracting policies will directly influence the drift. 
\end {enumerate}
\end{rem}
It is well known that for any Lebesgue measurable  control $u_1(\cdot)$ and $u_2(\cdot)$, taking values on compact sets $[0,  \bar{u}_1]$ and $[\underbar{u}_2,  \bar{u}_2]$ respectively,  the equation (\ref{model}) has a unique solution, such control processes will be called admissible controls. For more, one can refer to  $\O$ksendal and Sulem \cite{oksendal}. For each initial data $(x,i)$ we denote by ${\cal U}_j(x,i)$ the set of admissible controls which is just the set of all controls $u_j(\cdot)$ taking values in $U_j$ such that $X(0)=x$, $\al(0) = i$  and which are $\{{\cal F}_t\}_{t\geq 0}$-adapted where $ {\cal F}_t = \sigma\{\alpha(\xi),\eta(\xi), W(\xi); \xi\leq t\}$, $j=1,2$. \\

Let $C( u_1)$ be  the extraction cost function, we assume that this function depends only  on the extraction rate $u_1$.  In modeling the extraction cost function, we assume naturally that the extraction cost function is proportional to the production function. Thus we will use production function models to specify the extraction cost function. One of the popular production function is the quadratic  production function. This is due to its simplicity and the fact that it captures most stages of the production cycle. Therefore we will assume that our cost function $C(u_1)$ has the same form as a quadratic production function.
  Thus, we set 
 \bea
C(u_1) = au_1^2, \quad a>0.
\eea
 For more on production function models one can refer to  Houthakker \cite{houthakker}, Filipe and Adams \cite{filipeadams}, Humphrey \cite{humphrey}, Mankin \cite{mankin} and Taylor \cite{Taylor}.\\
The total profit rate for operating the mine is
 \bea
 P(x, u_1)=x u_1 -C( u_1).
\eea
The mining company  pre-tax profit rate function  is $ \theta P(x, u_1)$. 
The government  profit rate without the tax revenue is $ (1-\theta)P(x, u_1)$.
 The total income tax the government levies on the mining company is $ u_2 \theta P(x, u_1)$.
 The post-tax profit rate of the company is 
 \bea
\ad L_1(x,u_1, u_2)= \theta P(x, u_1 )(1- u_2), 
\eea
and the government profit rate function is
 \bea
\ad L_2(x,u_1, u_2)= (1-\theta) P(x, u_1 ) +  u_2\theta P(x, u_1 ).
\eea
 Given a discount rate $r> 0$, the payoff of each player $ j=1,2$ is defined as follows;
 \bea
\ad  J_j(x,i;u_1,u_2)  \\
\ad  = E \bigg[\int_0^\infty e^{-rt}L_j(X(t),Y(t),u_1(t),u_2(t))\rm{d}t \bigg{|} X(0) = x, \al(0) = i \bigg].
\eea
 It is obvious each player wants to maximize its own payoff.
The company will try to maximize its payoff by adjusting the extraction rate $ u_1(\cdot)$, while the government will maximize its payoff by changing the tax rate $ u_2(\cdot)$ following the changes in the commodity price $ X(t)$. This is therefore the setting of a noncooperative game.   Our goal is to find a noncooperative { Nash equilibrium} $ (u_1^*, u_2^*)$  such that  
  \beq{equilibrium}
\ad  J_1(x,i;u_1^*,u_2^*)\geq J_1(x,i;u_1,u_2^*), \\
\ad \hbox{ for all }u_1(\cdot) \in {\cal U}_1(x,i),\nonumber \\
\ad  J_2(x,i;u_1^*,u_2^*)\geq J_2(x,i;u_1^*,u_2),\\
\ad \hbox{ for all }u_2(\cdot) \in {\cal U}_2(x,i).\nonumber
\eeq
\section{Nash Equilibrium}
  \begin{defn}
Let $(u_1^*, u_2^*)$ be  a Nash equilibrium of our differential game, the functions 
  \beq{val1}
 V_1(x,i) = \sup_{u_1\in{\cal  U}_1}J_1(x,i; u_1,u_2^*) \nonumber \\
  V_2(x,i) = \sup_{u_2\in{\cal  U}_2}J_2(x,i; u_1^*,u_2) \nonumber
\eeq
are called value functions of Player 1 and Player 2 respectively.
\end{defn}
  In order  to find the optimal strategies $ u_1^*$ and $ u_2^*$ of the Nash equilibrium we first have to derive the value functions $ V_1$ and $ V_2$ of the differential game, then derive the optimal strategies. 
  Formally the value functions $ V_1$ and $V_2$ should satisfy  the Hamilton Jacobi Isaacs equations.
  Assuming that we have a Nash equilibrium $ u_1^*$ and $ u_2^*$ let us define corresponding Hamiltonians:
  \beq{Hamilton1}
&& H_1(x,i,V,\frac{\partial V}{\partial x}, \frac{\partial^2 V}{\partial x^2} \bigg)     \nonumber\\
&&=rV - \sup_{u_1\in U_1}  \bigg{(}\frac{1}{2}x^2\sigma^2(i)\frac{\partial^2V}{\partial x^2}  
+(x\mu(i) -\rho u_1)\frac{\partial V}{\partial x} \nonumber \\
&& + \nonumber  \int_{\rr} \bigg(V(x+\gamma(i)zx,i) -V(x,i)  -{\bf 1}_{\{|z|<1\}}(z)\frac{\partial V}{\partial x} \gamma(i)xz \bigg)\nu(dz) 
\disp  \\
&& + L_1(x,u_1,u_2^*)+  QV(x,\cdot)(i) \bigg{)},
\eeq	 
  and
  \beq{Hamilton2}
&& H_2(x,i,V,\frac{\partial V}{\partial x},\frac{\partial V}{\partial y}, \frac{\partial^2 V}{\partial x^2} \bigg)     \nonumber\\
&&=rV  - \sup_{u_2\in U_2}  \bigg{(}\frac{1}{2}x^2\sigma^2(i)\frac{\partial^2V}{\partial x^2} 
+(x\mu(i) -\rho u_1^*)\frac{\partial V}{\partial x}  \nonumber \\
&& + \nonumber  \int_{\rr} \bigg(V(x+\gamma(i)xz,y,i) -V(x,i)-{\bf 1}_{\{|z|<1\}}(z)\frac{\partial V}{\partial x} \gamma(i)zx \bigg)\nu(dz) 
\disp  \nonumber\\
&& + L_2(x,u_1^*,u_2)+  QV(x,\cdot)(i) \bigg{)},
\eeq	 
where 
 \bea
QV(x,\cdot)(i) = \sum_{j\neq i }q_{ij}[V(x,j) - V(x,i)].
\eea
  The corresponding Hamilton Jacobi Isaacs equations of this noncooperative game are
  \beq{isaac1}
 \left \{   \begin{array}{ll}   \disp H_1 \bigg(s,x,i,V_1,\frac{\partial V_1}{\partial x}, \frac{\partial^2 V_1}{\partial x^2} \bigg)=0&\\
 \disp   H_2 \bigg(s,x,i,V_2,\frac{\partial V_2}{\partial x}, \frac{\partial^2 V_2}{\partial x^2} \bigg)=0.&\\
\end{array}\right.
\eeq
 We define $ \tilde{\mu}_1 :=   (x\mu(i) -\rho u_1)$,  $ \tilde{L}_1 :=  L_1(x,u_1,u^*_2)$, \,\, 
$ \tilde{\mu}_2 := (x\mu(i) -\rho u_1^*) $,  $ \tilde{L}_2 :=  L_2(x,u_1^*,u_2)$, we have the following result.
  \begin{thm}\label{NashEquilibrium}
Assume that there exists $ (u_1^*, u_2^*)\in {\cal U}_1\times {\cal  U}_2$ such that 
the nonlinear Hamilton Jacobi Isaacs equations (\ref{isaac1})  have classical solutions $ V_j(x,i)$, $ j=1,2$,
   \beq{optCont}
  \disp  u_j^* = \arg\max \bigg(  \tilde{\mu}_j\frac{\partial V_j}{\partial x}
 \disp  
+\tilde{ L}_j   \bigg), \quad j=1,2.  
\eeq
Then the pair $  (u^*_1, u^*_2)$ is a Nash equilibrium solution and 
$  J_j(x,i;u_1^*,u_2^*)  = V_j(x,i)$, $j=1,2$.
\end{thm}
 \paragraph{Proof}
 The proof relies on the fact that this problem can be  uncoupled and solved as  an optimal control problem.  In fact, if we  already have  $ u^*_2(\cdot)$, then the differential game problem becomes an optimal control problem with the only control variable $u_1(\cdot)$ the HJB equation of this new control problem is in fact
  \beq{hjb1}
  H_1 \bigg(x,i,W_1,\frac{\partial W_1}{\partial x}, \frac{\partial^2 W_1}{\partial x^2} \bigg)=0.
\eeq
Following the assumptions of the Theorem, it is clear that the HJB equation (\ref{hjb1}) has a solution $V_1$ and the optimal policy of this new control problem is $u_1^*$. Therefore $u_1^*$ is in equilibrium with $u_2^*$ and $V_1$ is the value function of  Player 1. A similar argument can be used to show that $u_2^*$ is in equilibrium with $u_1^*$ and that $V_2$ is the value function of  Player 2.
\endproof

\section{Closed-form Solutions}
The following theorem presents the main result of this paper.
  \begin{thm}\label{mainThm}
Assume  that there exists a Nash equilibrium $(u_1^*, u_2^*)$ such that $u_2^*$ is an open loop control, in other terms, $u_2^*$ does not depend on the variable $x$. Then the solutions of the Hamilton Jacobi Isaacs equations \eqref{isaac1} are 
  \beq{solutions}
 V_1(x,i) = A_1(i)x^2, \quad V_2 =  A_2(i)x^2, \quad i=1,2,...,m.
\eeq
Player 1 optimal strategy is 
  \begin{eqnarray}
u_1^*(x,i) &=& \bigg (\frac{1}{2a}-\frac{\rho A_1(i)}{a\theta(1-u_2^*(i))} \bigg)x, \label{control1}
\end{eqnarray}
thus $u_1^*(x,i)$ can be expressed as $u_1^*(x,i) = K(i) x$.\\
Player 2 optimal  strategy is
  \beq{control2}
u_2^*(i) &=& \left \{   \begin{array}{ll}   \bar{u}_2 &\hbox{if}\quad K(i)-aK(i)^2 >0,\\
\underbar{u}_2 &\hbox{if}\quad K(i)-aK(i)^2 <0.
\end{array}
\right.
\eeq
Moreover, the coefficients $A_1(i), i=1,2,...,m$ satisfy the system of quadratic equations
  \beq{quad}
&&A_1(i)^2 \frac{\rho^2 }{a\theta(1-u_2^*(i))}+ A_1(i) \bigg(\sigma(i)^2- r+2\mu(i)- \sum_{j\neq i}q_{ij}-\frac{\rho}{a}\nonumber\\
&&\int_{\rr} \big(2\gamma(i)z +\gamma(i)^2z^2   -{\bf 1}_{\{|z|<1\}}(z)2 \gamma(i)z \big)\nu(dz)  \bigg) \\
&&+ \frac{\theta(1-u_2^*(i))}{4a}+ \sum_{j\neq i}q_{ij}A_1(j)=0, \qquad i=1,2,....,m, \nonumber
\eeq
and the coefficients $A_2(i), i=1,2,...,m$ satisfy the following linear system
  \beq{linear}
0&=&A_2(i) \bigg( \sigma(i)^2-r  
  +2(\mu(i) -\rho K(i))  - \sum_{j\neq i}q_{ij} \nonumber\nonumber \\
&&+  \int_{\rr} \big(2\gamma(i)z+ \gamma(i)^2z^2  -{\bf 1}_{\{|z|<1\}}(z)2 \gamma(i)z \big)\nu(dz) \bigg)   \\
&&\disp+ (K(i)-aK(i)^2)(1-\theta+\theta  \bar{u}_2{\bf 1}_{\{K(i)-aK(i)^2>0\}} \nonumber \\
&&    + \theta  \underbar{u}_2{\bf 1}_{\{K(i)-aK(i)^2<0\}} )+  \sum_{j\neq i}q_{ij}A_2(j), \quad \,\,\, i=1,2,...,m.\nonumber
\eeq
\end{thm}
\paragraph{Proof}
 In order to solve   \eqref{isaac1}, we will seek  solutions in the form
   \begin{eqnarray}
 V_1(x,i) = A_1(i)x^2, \quad V_2 =  A_2(i)x^2.  
 \end{eqnarray}
Therefore, we should have
  \beq{first}
0&=&rA_1(i)x^2 - \sup_{u_1\in U_1}  \bigg{[}x^2\sigma^2(i)A_1(i)  +2(x\mu(i) -\rho u_1) xA_1(i)
  \nonumber \\
&&  +   A_1(i)x^2\int_{\rr} \bigg((1+\gamma(i)z)^2-1   -{\bf 1}_{\{|z|<1\}}(z)2 \gamma(i)z \bigg)\nu(dz)  \\
&& + \theta(xu_1-au_1^2)(1-u_2^*)  +  \sum_{j\neq i}q_{ij}[(A_1(j)-A_1(i))x^2] \bigg{]},\nonumber
\eeq	 
and
  \beq{second}
0&=&rA_2(i)x^2  - \sup_{u_2\in U_2}  \bigg{[}x^2\sigma^2(i)A_2(i)  
  +2(x\mu(i) -\rho u_1^*) xA_2(i) + \nonumber\nonumber \\
&&  A_2(i)x^2\int_{\rr} \bigg((1+\gamma(i)z)^2-1  -{\bf 1}_{\{|z|<1\}}(z)2 \gamma(i)z \bigg)\nu(dz)   \\
&&+ (1-\theta+u_2\theta)(xu_1^*-a(u_1^*)^2)  +  \sum_{j\neq i}q_{ij}[(A_2(j)-A_2(i))x^2] \bigg{]}.\nonumber
\eeq	 
For the optimality, it is necessary to have
   \begin{eqnarray}
-2\rho  x A_1(i) +\theta(x-2a u_1)(1-u_2^*)&=&0.\label{firstreq}
 \end{eqnarray}
Therefore we should set
  \beq{Ai}
u_1^*(x,i) = \bigg (\frac{1}{2a}-\frac{\rho A_1(i)}{a\theta(1-u_2^*)} \bigg)x.
\eeq
Moreover, given that the operator $H_2$ is a linear function $u_2$, thus the control $u_2(\cdot)$ will necessary be a bang-bang control depending on the sign of the quantity 
  \beq{ee}
\theta(x (u^*_1)-a(u_1^*)^2)=\theta(K(i)-aK(i)^2)x^2,
\eeq
where $K$ is such that $u_1^*(x,i) = K(i)x$. It is clear that when $u_1^* $ is specified then $u_2^*$ can be obtained as follows 
  \beq{ee2}
u^*_2(x,i) =\left\{  \begin{array}{ll}   \bar{u}_2 & \hbox{if }\,\,\,K(i)-aK(i)^2 >0\\
               \underbar{u}_2  & \hbox{if }\,\,\,K(i)-aK(i)^2<0.
\end{array}
\right.
\eeq
It comes from \eqref{ee2} that $u_2^*$ is independent of the variable of $x$, so $u_2^*(x,i)\equiv u_2^*(i)$.\\
Plugging $u_1^*$ in \eqref{first} we obtain
  \beq{firstly}
0&=&-r A_1(i)x^2  + x^2\sigma^2(i)A_1(i) + x^2 \sum_{j\neq i}q_{ij}[(A_1(j)-A_1(i))] \nonumber \\
&&  +2x^2 \bigg(\mu(i) -\rho   \big(\frac{1}{2a}-\frac{\rho A_1(i)}{a\theta(1-u_2^*)} \big) \bigg) A_1(i)
  \nonumber \\
&&  +   A_1(i)x^2\int_{\rr} \bigg((1+\gamma(i)z)^2-1   -{\bf 1}_{\{|z|<1\}}(z)2 \gamma(i)z \bigg)\nu(dz)  \\
&& + x^2\theta   \big(\frac{1}{2a}-\frac{\rho A_1(i)}{a\theta(1-u_2^*)} \big)  ig[1-a   \big(\frac{1}{2a}-\frac{\rho A_1(i)}{a\theta(1-u_2^*)} \big) ig](1-u_2^*).    \nonumber
\eeq
After simplifying \eqref{firstly} we get
  \beq{new_e}
0&=&-r A_1(i)  + \sigma^2(i)A_1(i)  + \sum_{j\neq i}q_{ij}(A_1(j)-A_1(i))\nonumber \\
&&  +2\mu(i)A_1(i) -\frac{\rho A_1(i)}{a}+\frac{2\rho^2 A_1(i)^2}{a\theta(1-u_2^*)}
  \nonumber \\
&&  +   A_1(i)\int_{\rr} \bigg((1+\gamma(i)z)^2-1   -{\bf 1}_{\{|z|<1\}}(z)2 \gamma(i)z \bigg)\nu(dz) \nonumber \\
&& + \frac{\theta(1-u_2^*)}{4a}-\frac{\rho^2  A_1(i)^2}{a\theta(1-u_2^*)},    \nonumber\\
0&=&A_1(i)^2 \frac{\rho^2 }{a\theta(1-u_2^*)}+ A_1(i) \bigg(\sigma(i)^2- r+2\mu(i)- \sum_{j\neq i}q_{ij}-\frac{\rho}{a}\nonumber\\
&&\int_{\rr} \big(2\gamma(i)z +\gamma(i)^2z^2   -{\bf 1}_{\{|z|<1\}}(z)2 \gamma(i)z \big)\nu(dz)  \bigg) \\
&&+ \frac{\theta(1-u_2^*)}{4a}+ \sum_{j\neq i}q_{ij}A_1(j), \qquad i=1,2,...,m. \nonumber
\eeq
Similarly, given $u_1^*$ and using the fact from \eqref{Ai} that $u_1^*$ is of the form $u^*(x,i) = K(i)x$ and plugging the corresponding expression of $u_2^*(i)$ from \eqref{ee2} in \eqref{second} we get
  \beq{new_b}
0&=&-rA_2(i)x^2  + x^2\sigma^2(i)A_2(i)  
  +2(\mu(i) -\rho K(i)) x^2A_2(i) + \nonumber\nonumber \\
&&  A_2(i)x^2\int_{\rr} \bigg((1+\gamma(i)z)^2-1  -{\bf 1}_{\{|z|<1\}}(z)2 \gamma(i)z \bigg)\nu(dz)\nonumber   \\
&&\disp+x^2(K(i)-aK(i)^2) (1-\theta+\theta  \bar{u}_2{\bf 1}_{\{K(i)-aK(i)^2>0\}}\\
&& +  \theta  \underbar{u}_2{\bf 1}_{\{K(i)-aK(i)^2<0\}})+x^2 \sum_{j\neq i}q_{ij}(A_2(j)-A_2(i)).\nonumber
\eeq 
After simplifying \eqref{new_b} we obtain
  \beq{last_second}
0&=&A_2(i) \bigg( \sigma(i)^2-r  
  +2(\mu(i) -\rho K(i))  - \sum_{j\neq i}q_{ij} \nonumber\nonumber \\
&&+  \int_{\rr} \big(2\gamma(i)z+ \gamma(i)^2z^2  -{\bf 1}_{\{|z|<1\}}(z)2 \gamma(i)z \big)\nu(dz) \bigg)   \\
&&\disp+ (K(i)-aK(i)^2)(1-\theta+\theta  \bar{u}_2{\bf 1}_{\{K(i)-aK(i)^2>0\}}   \nonumber \\
&&     \theta  \underbar{u}_2{\bf 1}_{\{K(i)-aK(i)^2<0\}}   )+  \sum_{j\neq i}q_{ij}A_2(j),\,\,\, i=1,2,...,m.\nonumber
\eeq
In fact, \eqref{last_second} is a system of linear equations. 
\endproof

  \begin{rem} Let us set
  \beq{Peq}
P(i)&&\disp =  \sigma(i)^2-r  
  +2(\mu(i) -\rho K(i))  - \sum_{j\neq i}q_{ij}  \nonumber\\
&&\disp+  \int_{\rr} \big(2\gamma(i)z+ \gamma(i)^2z^2  -{\bf 1}_{\{|z|<1\}}(z)2 \gamma(i)z \big)\nu(dz), \quad i=1,2,...,m,
\eeq
and 
  \beq{Req}
R(i) \disp &=&-(K(i)-aK(i)^2)(1-\theta+\theta  \bar{u}_2{\bf 1}_{\{K(i)-aK(i)^2>0\}}    \nonumber \\
&&  + \theta  \underbar{u}_2{\bf 1}_{\{K(i)-aK(i)^2<0\}} ), \quad i=1,2,....,m.
\eeq
 In the particular case of a two-state Markov chain, the system \eqref{last_second} becomes
  \bea
 \left\{   \begin{array}{ll}  P(1) A_2(1)+q_{12} A_2(2)=  R(1),\\
                                    q_{21}A_2(1) +  P(2) A_2(2)=R(2),
\end{array}
\right.
 \eea
 and the solutions are 
   \begin{eqnarray}
 A_2(1) &&\disp= \frac{R(1)P(2)-q_{12}R(2)}{P(1)P(2)-q_{12}q_{21}}, \label{A2sol1}\\
 A_2(2)&&\disp =\frac{P(1)R(2)-q_{21}R(1)}{P(1)P(2)-q_{12}q_{21}}\label{A2sol2}.
 \end{eqnarray}
\end{rem}
Now, let us study the problem at hand in the particular case where the L\'{e}vy measure follows an exponential distribution.  We assume that the L\'{e}vy measure is of the form
 \bea
\nu(dz)  =\left\{  \begin{array}{ll} \eta e^{-\eta z} dz & \hbox{ if }\,\, z\geq 0,\\
                                             0  & \hbox{ if } \,\, z<0,
\end{array}
\right .
\quad $ for some $\quad  \eta>0.
\eea
Given that we have specified the form of the L\'{e}vy measure, we can now simplify \eqref{new_e} and \eqref{last_second}. Thus when the L\'{e}vy measure is exponential, the coefficients $A_1(i)$  should satisfy the quadratic equation
  \beq{expoLevy}
0&=&A_1(i)^2 \frac{\rho^2 }{a\theta(1-u_2^*)}+ A_1(i) \bigg(\sigma(i)^2- r+2\mu(i)- \sum_{j\neq i}q_{ij}-\frac{\rho}{a}\nonumber\\
&&\int_0^\infty \big(2\gamma(i)z +\gamma(i)^2z^2   -{\bf 1}_{\{|z|<1\}}(z)2 \gamma(i)z \big)\eta e^{-\eta z}dz  \bigg) \\
&&+ \frac{\theta(1-u_2^*)}{4a}+ \sum_{j\neq i}q_{ij}A_1(j) \nonumber\\
0&=&A_1(i)^2 \frac{\rho^2 }{a\theta(1-u_2^*)}+ A_1(i) \bigg(\sigma(i)^2- r+2\mu(i)- \sum_{j\neq i}q_{ij}-\frac{\rho}{a}\nonumber\\
&&+2\gamma(i)\frac{\gamma(i)+(1+\eta)\eta e^{-\eta}}{\eta^2}  \bigg)+ \frac{\theta(1-u_2^*)}{4a}+ \sum_{j\neq i}q_{ij}A_1(j), \,\,\, i=1,2.  
\eeq
Moreover, \eqref{Peq}  can be simplified as  follows
 \bea
P(i) & \disp =  \sigma(i)^2-r  
  +2(\mu(i) -\rho K(i))  - \sum_{j\neq i}q_{ij}  + 2\gamma(i)\frac{\gamma(i)+(1+\eta)\eta e^{-\eta}}{\eta^2},\,\, i=1,2,
\eea
We summarize these findings in the following corollary.
  \begin{cor}
If the L\'{e}vy process has finite activity and the L\'{e}vy measure has the form $\nu(dz) = \eta e^{-\eta z} dz, z>0$, for some $\eta>0$,  the value functions and the optimal policies obtained in \thmref{mainThm} are such that the coefficients $A_1(i)$ satisfy the system of quadratic equations
  \beq{cor1_1}
&&A_1(i)^2 \frac{\rho^2 }{a\theta(1-u_2^*)}+ A_1(i) \bigg(\sigma(i)^2- r+2\mu(i)- \sum_{j\neq i}q_{ij}-\frac{\rho}{a}\nonumber\\
&&+2\gamma(i)\frac{\gamma(i)+(1+\eta)\eta e^{-\eta}}{\eta^2}  \bigg)+ \frac{\theta(1-u_2^*)}{4a}+ \sum_{j\neq i}q_{ij}A_1(j)=0, \quad i=1,2,...m,  
\eeq
and in the particular case of a two-state Markov chain, the coefficients $A_2(i)$ are defined as follows
 \bea
 A_2(1) &\disp= \frac{R(1)P(2)-q_{12}R(2)}{P(1)P(2)-q_{12}q_{21}}, \\
 A_2(2)&\disp =\frac{P(1)R(2)-q_{21}R(1)}{P(1)P(2)-q_{12}q_{21}}
\eea
with 
 \bea
P(i)& \disp =  \sigma(i)^2-r  
  +2(\mu(i) -\rho K(i))  - \sum_{j\neq i}q_{ij}  + 2\gamma(i)\frac{\gamma(i)+(1+\eta)\eta e^{-\eta}}{\eta^2},\,\, i=1,2,\\
  R(i) &\disp =-\big(1-\theta+\theta  ( \bar{u}_2{\bf 1}_{\{K(i)-aK(i)^2>0\}}+ \underbar{u}_2{\bf 1}_{\{K(i)-aK(i)^2<0\}})\big)(K(i)-aK(i)^2), \quad i=1,2.
\eea
\end{cor}

It is well known through empirical evidences that commodity prices follow L\'{e}vy processes with infinite jumps activity. For more on this finding one can refer to \cite{aitsahalia} and the references therein.  It is therefore interesting that we cover that aspect of this problem. In this section, we derive optimal extraction policies when the L\'{e}vy measure is given by 
  \beq{levy}
\nu(dz) =\left\{  \begin{array}{ll}\disp  \frac{e^{-|z|}}{|z|^2} dz &  \hbox{ if }\,\, z\neq 0,\\
                                             0  &  \hbox{ if }\,\, z=0 . 
\end{array}
\right.
\eeq
It can be shown easily that the measure defined in \eqref{levy} satisfies the L\'{e}vy-Khintchine formula $\disp \int_\rr \min(|z|^2,1) \nu(dz)<\infty$  and that $\disp \int_\rr \nu(dz)=\infty$.
It comes that the coefficients $A_1(i), i=1,2$ such satisfy the quadratic equations
  \beq{quad_2}
0&=&A_1(i)^2 \frac{\rho^2 }{a\theta(1-u_2^*)}+ A_1(i) \bigg(\sigma(i)^2- r+2\mu(i)- \sum_{j\neq i}q_{ij}-\frac{\rho}{a}\nonumber\\
&&\int_{\rr} \big(2\gamma(i)z +\gamma(i)^2z^2   -{\bf 1}_{\{|z|<1\}}(z)2 \gamma(i)z \big)\frac{e^{-z}}{|z|^2}dz  \bigg) \nonumber\\
&&+ \frac{\theta(1-u_2^*)}{4a}+ \sum_{j\neq i}q_{ij}A_1(j) \nonumber\\
&=&
A_1(i)^2 \frac{\rho^2 }{a\theta(1-u_2^*)}+ A_1(i) \bigg(\sigma(i)^2- r+2\mu(i)- \sum_{j\neq i}q_{ij}-\frac{\rho}{a}\nonumber\\
&&\int_{\rr}2\gamma(i)z\frac{e^{-|z|}}{z^2}dz  +\int_\rr \gamma(i)^2z^2\frac{e^{-|z|}}{z^2}dz    -\int_{-1}^1 2 \gamma(i)z\frac{e^{-|z|}}{z^2}dz  \bigg) \nonumber\\
&&+ \frac{\theta(1-u_2^*)}{4a}+ \sum_{j\neq i}q_{ij}A_1(j) 
\eeq
Consequently we get,
  \beq{vad_2}
&&A_1(i)^2 \frac{\rho^2 }{a\theta(1-u_2^*)}+ A_1(i) \bigg(\sigma(i)^2- r+2\mu(i)- \sum_{j\neq i}q_{ij}-\frac{\rho}{a}\nonumber\\
&&\int_{-\infty}^\infty 2\gamma(i)\frac{e^{-|z|}}{z}dz  +2\int_0^\infty \gamma(i)^2e^{-|z|}dz    -\int_{-1}^1 2 \gamma(i)\frac{e^{-|z|}}{z}dz  \bigg) \nonumber\\
&&+ \frac{\theta(1-u_2^*)}{4a}+ \sum_{j\neq i}q_{ij}A_1(j)=0.   \nonumber
\eeq
Which simplifies to
  \beq{levy25}
&&A_1(i)^2 \frac{\rho^2 }{a\theta(1-u_2^*)}+ A_1(i) \bigg(\sigma(i)^2- r+2\mu(i)- \sum_{j\neq i}q_{ij}-\frac{\rho}{a}\nonumber\\
&&  +2 \gamma(i)^2   \bigg) + \frac{\theta(1-u_2^*)}{4a}+ \sum_{j\neq i}q_{ij}A_1(j) =0,\quad i=1,2,...,m.  
\eeq
In addition, \eqref{Peq} becomes
 \bea
P(i) \disp= \sigma(i)^2-r  
  +2(\mu(i) -\rho K(i))  - \sum_{j\neq i}q_{ij}  \disp+ 2\gamma(i)^2, \quad i=1,2,...,m.
\eea
The following corollary summarizes this result.
  \begin{cor}
If the L\'{e}vy process has infinite activity and the L\'{e}vy measure has the form $\nu(dz) =  e^{-| z|}/|z|^2 dz, z\neq 0$, the value functions and the optimal policies obtained in \thmref{mainThm} are such that the coefficients $A_1(i)$ satisfy the system of quadratic equations
  \beq{lcor2-1}
&&A_1(i)^2 \frac{\rho^2 }{a\theta(1-u_2^*)}+ A_1(i) \bigg(\sigma(i)^2- r+2\mu(i)- \sum_{j\neq i}q_{ij}-\frac{\rho}{a}\nonumber\\
&&  +2 \gamma(i)^2   \bigg) + \frac{\theta(1-u_2^*)}{4a}+ \sum_{j\neq i}q_{ij}A_1(j) =0,\quad i=1,2,...m.  
\eeq
and  in the particular case of a two-state Markov chain, the coefficients $A_2(i)$ are defined as follows
  \begin{eqnarray}
 A_2(1) &\disp=& \frac{R(1)P(2)-q_{12}R(2)}{P(1)P(2)-q_{12}q_{21}}, \label{infA1}\\
 A_2(2)&\disp =&\frac{P(1)R(2)-q_{21}R(1)}{P(1)P(2)-q_{12}q_{21}}\label{infA2}
\end{eqnarray}
with 
 \bea
P(i)& \disp = \sigma(i)^2-r  
  +2(\mu(i) -\rho K(i))  - \sum_{j\neq i}q_{ij}  \disp+ 2\gamma(i)^2, \quad i=1,2,\\
  R(i) &\disp =-\big(1-\theta+\theta  ( \bar{u}_2{\bf 1}_{\{K(i)-aK(i)^2>0\}}+ \underbar{u}_2{\bf 1}_{\{K(i)-aK(i)^2<0\}})\big)(K(i)-aK(i)^2), \quad i=1,2.
\eea
\end{cor}



\section{Application}
In this section we  study two cases of the extraction and taxation problems, we will first analyze the case where the country is medium size produce of the commodity and in the second case we we analyze the problem when the country is a major producer of the commodity. In both case we assume that $\bar{u}_2=0.2$ and $\underbar{u}_2=0.$
\subsection{Medium  producer}
 Consider a Profit Sharing Agreement between a medium gold producer and a multinational mining company where the company takes { 30\%} and the country keeps {70\%} of the profits, thus $  \theta =0.3.$ 
 Because the country is a medium producer, the country production level will not influence the commodity price, thus $  \rho =0$ and the commodity price follows the following SDE
 \bea
  dX(t) = X(t) \bigg(\mu(\al(t))dt + \sigma(\al(t))dW_t + \int_\rr \lambda(\al(t))z  \bar{N}(dt,dz) \bigg).
\eea

 We assume that the gold market has two trends, $   {\cal M} =\{ 1,2 \}$, $  \al(t)=1 $ represents the uptrend and $  \al(t)=2$ represents the downtrend.
 Moreover we assume, $  r=0.02$, $  \mu=(0.08, -0.1)$, $  \sigma = (0.2, 0.3)$, $  \gamma= (0.05, 0.09),$ and $   \nu(dz) = 5 e^{- 5z} dz, z>0$. 
 The generator of the Markov chain $  \al(t)$ is 
$
\disp
  Q= \bigg(  \begin{array}{ll} -0.4 & 0.4 \\
             0.1 &-0.1 
              \end{array}  \bigg).
$
 The extraction cost function is $  \disp C(u ) = 15 u^2$, so $   a =15$. Note that $  u$ is in millions of ounces of gold per year, and the unit of the cost function $  C(u)$ is millions of dollars per year.
 If   $  u_2^*\equiv  \bar{u}_2=0.2$ then
$
  V_1(x,i) = A_1(i) x^2,\,\,\, i=1,2,
$
such that $  A_1(1)$ and $  A_2(2)$ solve the system 
 \bea
  \left\{   \begin{array}{ll}  -0.2190A_1(1) +0.4 A_1(2)+0.004 =0\\
 0.1 A_1(1)-0.2279A_1(2) +0.004 =0,
\end{array}
\right.
\eea
and the solutions are $  A_1(1)=0.2535$ and $  A_1(2)=0.1288$.\\
and the optimal extraction rates are $  u_1^*(x,1)=u_1^*(x,2)=   \frac{1}{30}x$.
If   $  u_2^*\equiv \underbar{u}_2=0$ then $  A_1(1)$ and $  A_1(2)$ solve a slightly different system
 \bea
  \left\{   \begin{array}{ll}  -0.2190A_1(1) +0.4 A_1(2)+ 0.0050 =0\\
 0.1 A_1(1)-0.2279A_1(2) + 0.0050 =0,
\end{array}
\right.
\eea
and $  A_1(1)=0.3169$ and $  A_1(2)=0.1610.$ 
 Given that $  u^*_1(x,1)=u_1^*(x,2)=\frac{1}{2a}x$, thus $  K(i)=\frac{1}{2a}$. It is clear that $  K(i)-aK(i)^2 = \frac{1}{2a} -\frac{1}{4a}>0$, thus $  u_2(1)=u_2(2)=  \bar{u}_2=0.2$.\\
 In sum the Nash Equilibrium is 
 \bea
  u_1^*(x,i)=\frac{1}{2a}x,\quad $ and $   \quad  u^*_2(x,i)=   \bar{u}_2=0.2.
\eea

\subsection{Major Producer}
Consider an oil company with an extraction lease of an oil field with a known reserve of $ M=10$ billion barrels. We assume that the  profit sharing agreement between the oil company and the government is such that the oil company takes 20\% of profits and the government takes  80\%, so $\theta=0.2.$
  The yearly discount rate $r=0.02$, the  yearly return vector is $\mu=(0.02,-0.1)$, the  yearly volatility vector is $\sigma=(0.2,0.3)$, the  yearly intensity vector is $\gamma=(0.022,0.03)$,  and the generator of the Markov chain is
  \bea
\disp
Q= \bigg(  \begin{array}{ll} -0.3 & 0.3 \\
             0.5 &-0.5 
              \end{array}  \bigg).
\eea
The parameter $ \rho\in[0,1)$ will capture the relative impact of  the oil production on the oil price, in this example $\rho=0.001$. The extraction cost function is $\disp C(u ) = 2 u^2$, so $ a =2$. Note that, in the cost function $C(u)$, the argument  $u$ is in millions of barrels per year, and the unit of the cost function $C(u)$ is millions of dollars per year. We assume that the L\'{e}vy measure has infinity intensity $\disp \nu(dz) = \frac{e^{-|z|}}{|z|^2}, z\neq0$.\\
If   $u_2^*\equiv  \bar{u}_2=0.2$ then
 \bea
&V_1(x,i) = A_1(i) x^2\\
\eea
such that $A_1(1)$ and $A_2(2)$ solve the system 
 \bea
\left\{   \begin{array}{ll}  3.125\times 10^{-6} A_1(1)^2-0.239532A_1(1) +0.3 A_1(2)+0.02 =0\\
 3.125\times 10^{-6} A_1(2)^2-0.6287A_1(2) +0.5 A_1(1)+0.02 =0.
\end{array}
\right.
\eea
The acceptable solutions are $A_1(1) = 37.2674$ and $A_1(2) = 29.6747$. In fact this system has four solution pairs $(37.2674,  29.6747)$, $(194.783,  155.06)$, $(76534.2 - 138910i, 201092 + 110575i)$ and $(76534.2 + 138910i,  201092 - 110575i)$, but only the solution $(37.267,  29.674)$ is such that $u_1^*(x,i)\geq 0$ for all $x$ and $i$. Therefore we have
 \bea
&\disp V_1(x,1) =  37.2674x^2,\quad V_1(x, 2) = 29.6747 x^2,\\
&\disp u_1^*(x,1) =0.133539x, \qquad  u_1^*(x,2) = 0.157267 x.
\eea
It is worth noting that the value function $V(x,i)$ is given in millions of dollars and the extraction rate $ u^*(x, i)$ is given in millions of barrels per year. In fact the daily optimal extraction rate is
 \bea
&\disp u_1^*(x),i) = \frac{1}{365} \big(\frac{1}{2a}-\frac{\rho A_1(i)}{a\theta(1-u_2^*(i))} \big)x,\\
& u_1(x,1) = 0.00036x \quad $millions of barrels per day, $\\
&  u_1(x,2) =0.00043x \quad $millions of barrels per day. $
\eea
If $u_2^*\equiv\underbar{u}_2= 0$, then the coefficients $A_1(1) $ and $A_1(2)$ solve the  system
  \bea
\left\{   \begin{array}{ll}  2.5\times 10^{-6} A_1(1)^2-0.239532A_1(1) +0.3 A_1(2)+0.025 =0\\
 2.5\times 10^{-6} A_1(2)^2-0.6287A_1(2) +0.5 A_1(1)+0.025 =0.
\end{array}
\right.
\eea
Then acceptable solution is $A_1(1) = 46.5843$ and $ A_1(2)= 37.0934$. The value function and optimal extraction rate are 
 \bea
&V_1(x, 1) =  46.5843 x^2\quad V_1(x,2) =  37.0934x^2\\
&u_1^*(x,1) = 0.133539 x, \quad u_1^*(x,2) = 0.157267 x.
\eea
The daily optimal extraction rates are
\bea
& u_1(x,1) = 0.00036x \quad $millions of barrels per day, $\\
&  u_1(x,2) =0.00043x \quad $millions of barrels per day. $
\eea
Likewise, if $u_1^*(x,i) = K(i)x$ with $K(1)= 0.133539$ and $K(2) = 0.157267 $, then the coefficient $A_2(1)$ and $A_2(2)$ obtained from \eqref{infA1} and \eqref{infA2} are 
 \bea
A_2(1) =195.654 \quad $and$ \quad A_2(1)=155.792.
\eea
Thus the value function is
 \bea
V_2(x,1)=195.654 x^2 \quad $and $ V_2(x,2) = 155.792 x^2.
\eea
The optimal tax rate is $u_2^*(x,1)=u^*_2(x,2)=  \bar{u}_2=20\%$ because $K(1)-aK(1)^2=0.0978738>0$ and  $K(2)-aK(2)^2=0.107801>0$.\\
In Figure 1, we represent the value function of player 1 (the mining company)  when the market is up and when the market is down. 
In Figure 2, we represent the value function of player 2 (the government)  when the market is up and when the market is down.
\section{Conclusion}
In this work, we study the optimal  extraction and taxation  of a strategic natural resource. Assuming that the price of the natural resource follows a regime switching L\'{e}vy process, we model this extraction and taxation problem as a noncooperative stochastic differential game between a multinational  mining company and the government of a resource-rich nation. we derive closed-form formulas for the value functions as well as the optimal extraction and taxation policies.  We end this paper by showing how our result can be easily applied in the optimal management of a major oil field and the efficient taxation of a multinational oil corporation.

  \begin{center}
  \begin{figure}
\vspace{-0cm}
\includegraphics[height=17cm,width=13cm]{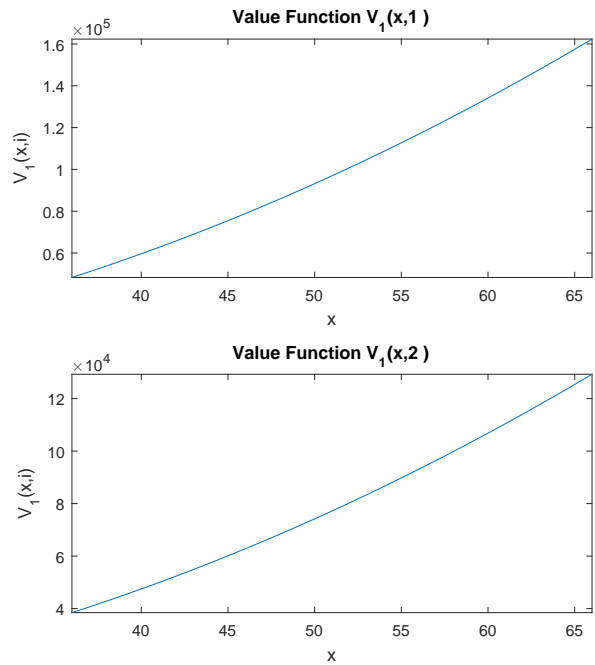}
\caption{This graph represents the value function of Player 1. }
\end{figure}
\end{center}
  \begin{center}
  \begin{figure}
\vspace{-0cm}
\includegraphics[height=17cm,width=13cm]{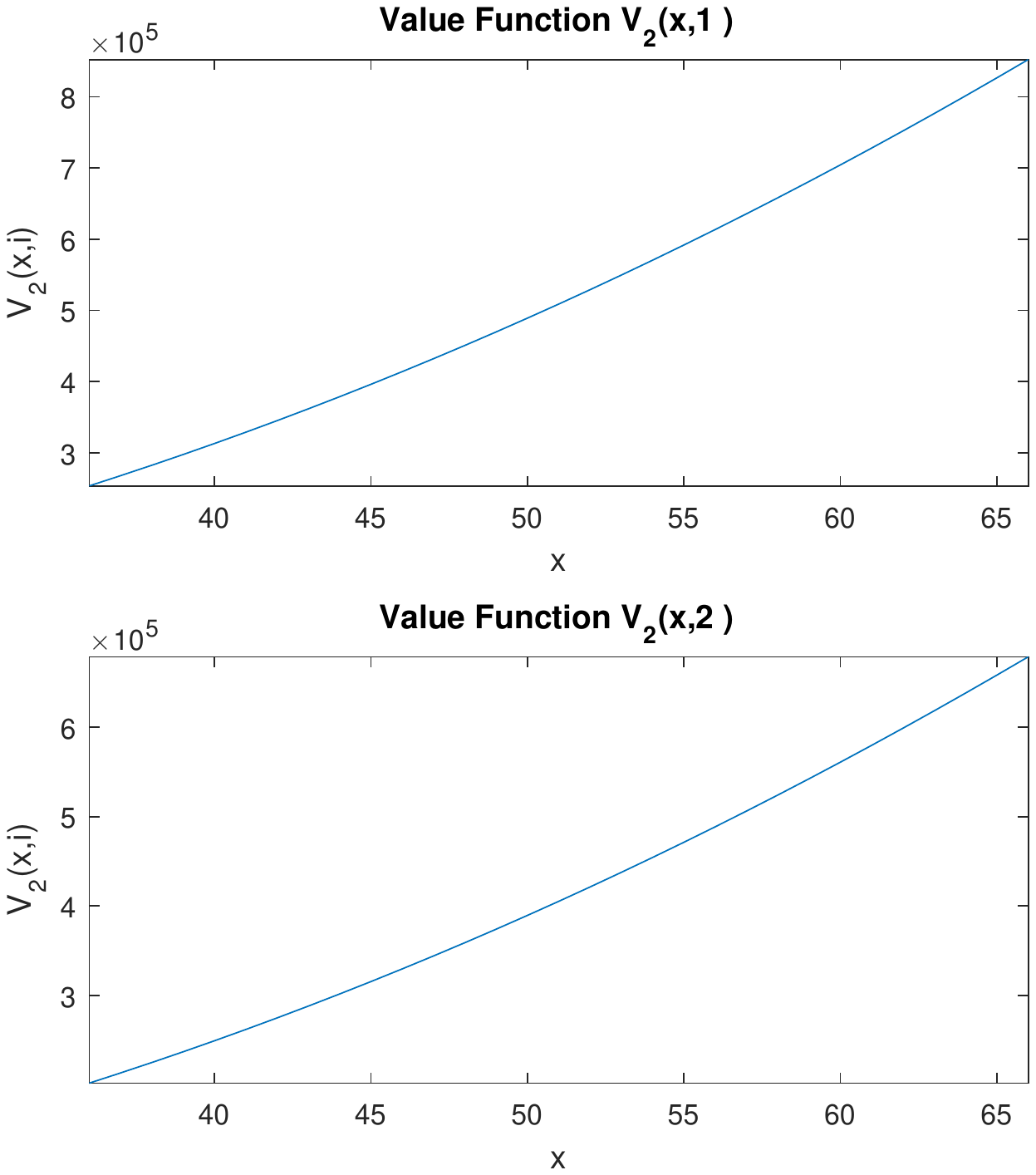}
\caption{This graph represents the value  function of Player 2. }
\end{figure}
\end{center}



\end{document}